\documentclass{elsart}
\usepackage{ifpdf}
\usepackage{graphicx,natbib,amssymb,lineno}
\usepackage{graphics}
\usepackage[%
  breaklinks=true,%
  colorlinks=true,%
  linkcolor=blue,anchorcolor=blue,%
  citecolor=blue,filecolor=blue,%
  menucolor=blue,pagecolor=blue,%
  urlcolor=blue]{hyperref}

\makeatletter
\def\elsartstyle{%
    \def\normalsize{\@setfontsize\normalsize\@xiipt{14.5}}
    \def\small{\@setfontsize\small\@xipt{13.6}}
    \let\footnotesize=\small
    \def\large{\@setfontsize\large\@xivpt{18}}
    \def\Large{\@setfontsize\Large\@xviipt{22}}
    \skip\@mpfootins = 18\p@ \@plus 2\p@
    \normalsize
}
\@ifundefined{square}{}{\let\Box\square}
\makeatother

\begin{document}

\begin{frontmatter}
\title{A beyond mean field analysis of the shape transition in the Neodymium isotopes.}
\author{Tom\'as R. Rodr\'iguez, J. Luis Egido}
\address{Departamento de F\'isica Te\'orica C-XI, Universidad Aut\'onoma de Madrid, 28049 Madrid, Spain}

\ead{tomas.rodriguez@uam.es,j.luis.egido@uam.es}
\begin{abstract}
The beyond mean field approximation combined with the Gogny interaction is used to analyze the spherical to prolate deformed shape transition in the Neodymium isotopic chain. The vibrator as well as the rotor limits are nicely reproduced  while the transitional region is only qualitatively described probably due
to the lack of triaxial correlations in the calculations. Our results do not support the interpretation of
$^{150}$Nd as a critical point nucleus and question the interpretation of shape changes as  nuclear shape phase transitions.
\end{abstract}
\end{frontmatter}


Phase transitions have always attracted the attention of scientist in all research fields. In nuclear physics, in particular, shape transitions and the super-fluid
normal-fluid transition have been extensively studied since a long time at finite temperature \cite{ER.JPG.93} in many different theoretical approaches as well as experimentally. Also at zero temperature the phenomena of shape coexistence and shape transition have been  a fascinating working field and many papers have been devoted to it \cite{A.Nature.00,IACHARI87}.
Most of these works have concentrated on the shape transitions from spherical to prolate
(oblate) or triaxial deformed shapes. In these studies there have been used several approaches  from the mean field approach to the Generator Coordinate Method
(GCM) with schematic interactions \cite{K.NPA.74} and  effective ones \cite{BENDRMF03} as well as 'exact' calculations with small configuration  spaces and realistic interactions \cite{PSF.NPA.00}.  However, the overwhelming contribution to this topic has been made within the framework of the interacting boson models (IBM) \cite{IACHARI87,SHOLANNP78,CAST79,GUPJPG95,IACZAMCASPRL98,GRABFPRC03}. In this algebraic approach, spherical, prolate and $\gamma$-soft nuclei are associated to the dynamical symmetries of the model, U(5), SU(3), O(6), respectively. Also the general aspects of the shape transitions have been described within the classical limit of the IBM which is the Bohr (collective) Hamiltonian \cite{IACHARI87}.\\ \indent
However, the concept of shape \textit{phase} transitions is, at least, controversial. In finite systems one does not expect real phase transitions because the large quantum fluctuations may spoil the sharp behavior of the corresponding parameters found in a macroscopic system leading to a blurred phase transition. In addition, the difficulties associated with the identification (non-uniqueness) of the order parameter as well as the lack of a continuous control parameter ($Z$, $N$ are integer numbers) have contributed to create a certain reluctance of part of the nuclear community with respect to this issue. 
With this background, some recent developments by Iachello \cite{IACHPRL00,IACHPRL01} came out as a surprise, since he went even further by proposing the existence of critical point nuclei in the nuclear shape phase transitions.
Iachello developed analytical solutions to the Bohr collective Hamiltonian with potentials which correspond, within the framework of the model, to the critical point of the first order shape phase transitions between spherical and prolate nuclei (the so-called X(5) symmetry), as well as to the critical point of the second order phase transition from  spherical to $\gamma$-soft nuclei (identified as the E(5) symmetry). 
Furthermore, these new solutions were used to assign a critical character to some nuclei which experimentally show similar excitation spectra as the ones predicted by these theoretical approaches. Hence, the isotopes $^{152}$Sm, $^{150}$Nd were associated to the X(5) symmetry \cite{CASTPRL01,KRUPRL02} while the nucleus $^{134}$Ba would correspond to the E(5) symmetry  \cite{CASTPRL00}.  \\ \indent
Although the IBM and the Bohr collective Hamiltonian are useful representations for describing shape phase transitions, it is important to mention that both models oversimplify the underlying nuclear many-body problem. Hence, the interacting boson models have many free parameters which are specifically adjusted to a particular region of nuclei and, therefore, their predictive power is limited. Concerning the collective Hamiltonian, one misses a compelling
justification of the potentials used in such approaches. For instance a derivation based on microscopic calculations with reasonable nucleon-nucleon interactions has not been presented yet. Furthermore, these potentials ignore nuclear degrees of freedom which might be significant in a transitional region and focus only on the quadrupole deformation. 

The aim of the present work is to analyze these shape transitions from a microscopic viewpoint using state of the art Beyond Mean Field Approaches (BMFA). In particular we will study the spherical to prolate deformed transition in the Neodymium isotopes ($Z=60$), with the intention, after comparison with
the experiment and the X(5) predictions,  to shed some light on this delicate problem of the nuclear phase transitions. 
There have been  some self-consistent mean field approaches with effective phenomenological interactions
which have also addressed the shape phase transitions in the $A\sim 150$ region  \cite{GOGPLB75,FOSBONATLALAPRC06,RS.07}. Recently,  Nik\u{s}i\'c and collaborators have reported the first BMFA study of the proposed critical behavior of the $^{150}$Nd isotope \cite{NVLR.PRL.07}. Their study is based on the Relativistic Mean Field approach and concentrates on the nuclei $^{148-150-152}$Nd. Our BMFA is based on the Hartree-Fock-Bogoliubov (HFB) approach and in the calculations we use the Gogny force. In this Letter we extend the calculations of
ref.~\cite{NVLR.PRL.07}  to describe the vibrational and rotational limits of the shape transitions, i.e. the nuclei $^{144-146-148-150-152-154}$Nd, and we furthermore analyze the relevance of triaxial shapes along the transition region.   In our BMFA \cite{RODPRL07} we  first generate the collective subspace of HFB  wave functions using $\vec{q}$-constrained particle number projection {\it before the variation}, i.e., we minimize the energy
\begin{eqnarray}
E^{N,Z}(\vec{q})=\frac{\langle{\Phi}^{N,Z}(\vec{q})|\hat{H}|{\Phi}^{N,Z}(\vec{q})\rangle}{\langle{\Phi}^{N,Z}(\vec{q})|{\Phi}^{N,Z}(\vec{q})\rangle},
\label{PNP-PES}
\end{eqnarray}
with $|\Phi^{N,Z}(\vec{q})\rangle=\hat{P}^{N}\hat{P}^{Z}|\phi(\vec{q})\rangle$, 
where $\hat{P}^{N(Z)}$ is the projector onto neutron (proton) number, $|\phi(\vec{q})\rangle$ 
are HFB-type wave functions and $\vec{q}\equiv q_{1},...,q_{M}$ is a set of intrinsic constraints that allows the definition of potential energy surfaces along the most relevant degrees of freedom. The configuration mixing calculation is performed within the generator coordinate method (GCM) framework taking linear combinations of the  particle number and {\it angular momentum projected} wave functions obtained in the first step
\begin{eqnarray}
|\Psi^{N,Z,J,\sigma}\rangle=\int f^{N,Z,J,\sigma}(\vec{q}) \; \hat{P}^{J}\;|{\Phi}^{N,Z}(\vec{q})\rangle d\vec{q}.
\label{GCMstate}
\end{eqnarray} 
Then, the variational principle applied to the weights $ f^{N,Z,J,\sigma}(\vec{q})$ gives the generalized eigenvalue problem (Hill-Wheeler equation):
\begin{eqnarray}
\int \left(\mathcal{H}^{N,Z,J}(\vec{q},\vec{q}^\prime)-E^{N,Z,J,\sigma}\mathcal{N}^{N,Z,J}(\vec{q},\vec{q}^\prime )\right)f^{N,Z,J,\sigma}(\vec{q}^\prime)d\vec{q}^\prime=0,
\label{HWeq}
\end{eqnarray}
with  $\mathcal{H}^{N,Z,J}$ and $\mathcal{N}^{N,Z,J}$ the norm and Hamiltonian overlaps, respectively, see \cite{RODGUZNPA02} for further details. We use in our calculations the finite range density dependent Gogny force
with the D1S parametrization \cite{BERGNPA84}, which is well known for its successful predictions.

\begin{figure} [t]
\centering
\includegraphics[angle=0,scale=0.5]{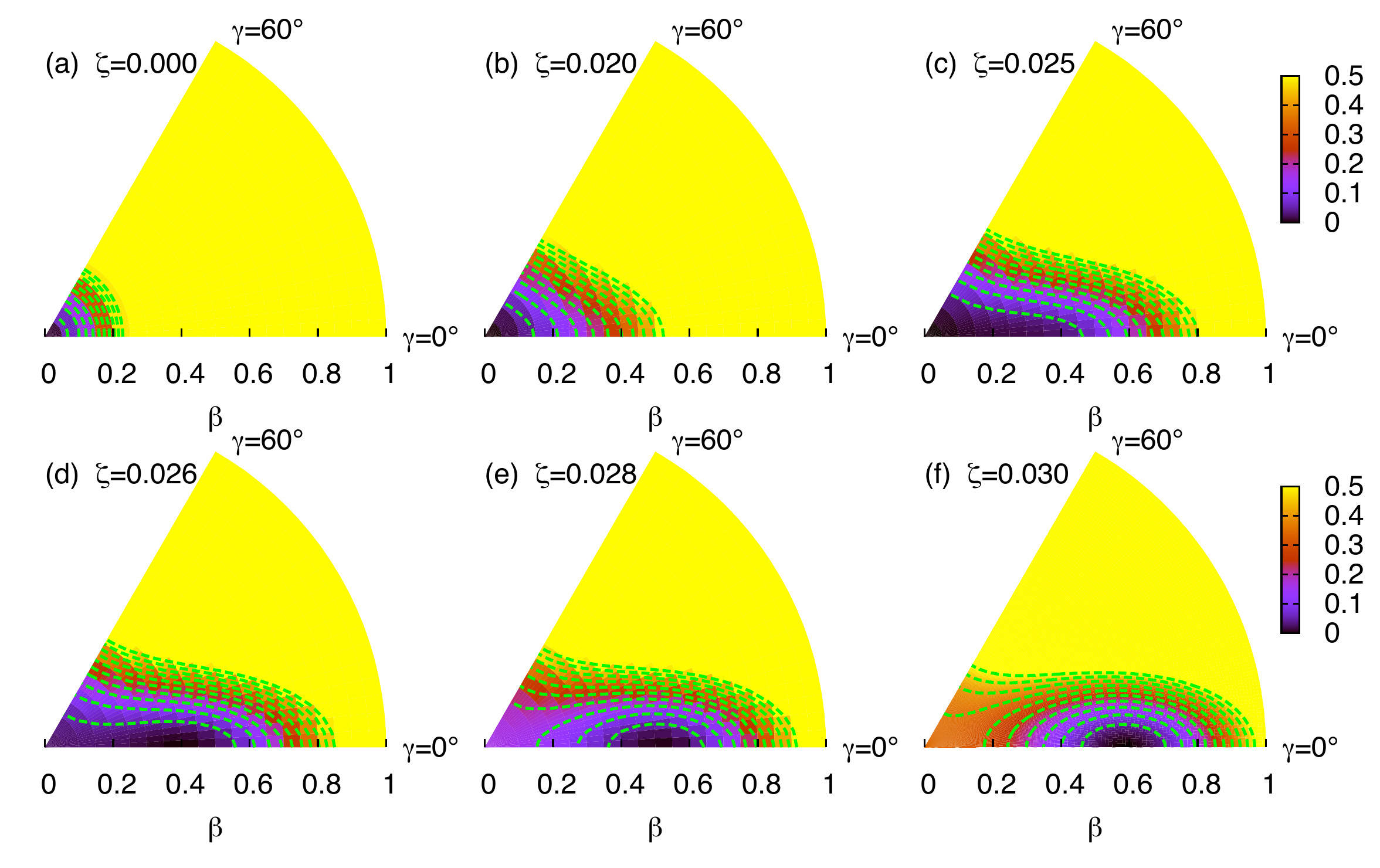} 
\caption{(a)-(f) Potential energy $V(\beta,\gamma)$ (Eq. \ref{bohr}) given by the collective model to describe the U(5)-SU(3) transition for different values of the control parameter $\zeta$. Contour lines are plotted each 0.05 arbitrary units, the energy is set to zero at the minimum of each surface.}
\label{Fig1}
\end{figure}
Let us now turn to the IBM description. The Bohr Hamiltonian allows the description of the quantum many-body system in terms of a one-body Hamiltonian with a potential energy $V$ which depends on the collective variables $(\beta,\gamma)$ \cite{IACHARI87,RINGSHUCK}, i.e. $H_{coll}= T(\beta,\gamma) +V(\beta,\gamma)$. The collective model can be solved analytically under certain symmetry restrictions and the comparison of these solutions with the experimental data is used to identify the intrinsic shape of a particular nucleus.  The potential energy surfaces (PES) $V(\beta,\gamma)$ which describe the spherical to prolate deformed transition (U(5)$\rightarrow$SU(3)) can be obtained in the framework of the interacting boson model using the method of intrinsic states \cite{IACHPRL01}:
\begin{equation}
\label{bohr}
V(\beta,\gamma)=\frac{N\beta^{2}}{1+\beta^{2}}\left[1+\frac{5}{4}\zeta\right]-\frac{N(N-1)}{(1+\beta^{2})^{2}}\zeta\left[4\beta^{2}+2\sqrt{2}\beta^{3}\mathrm{cos} 3\gamma+\frac{1}{2}\beta^{4}\right],
\end{equation}
where $N$ is the number of valence pairs (we take $N=10$ as in Ref.~\cite{IACHPRL01}) and $\zeta$, ($0\leq\zeta\leq1$), is the continuous control parameter which determines the final structure of the potential and is used to simulate the phase transition. In Fig.~\ref{Fig1} we show this potential, in the  $(\beta,\gamma)$ plane, for the most significant values of $\zeta$. For $\zeta =0$, see panel (a), we
find a potential very stiff in the $\beta$, $\gamma$ variables which correspond to a spherical nucleus. For higher $\zeta$ values (panels (b), (c) and (d)) we find a  large softening in the
$\beta$ deformation and a smaller one in the $\gamma$ direction. For larger $\zeta$ values (panels (e) and (f)) the nucleus becomes prolate deformed and rather stiff in the $\gamma$ direction. 
\begin{figure} [t]
\centering
\includegraphics[angle=0,scale=0.5]{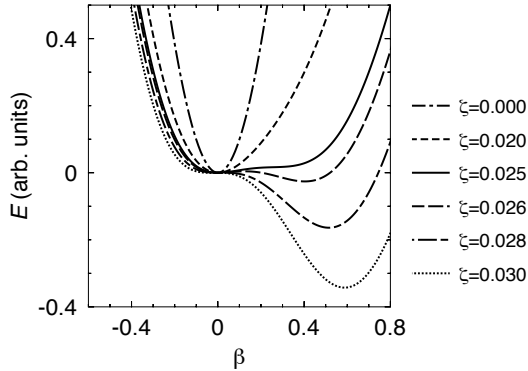} 
\caption{Potential energy (Eq. \ref{bohr}) without normalization along the axial trajectory $\gamma=0^{\circ},60^{\circ}$.}
\label{Fig2}
\end{figure}
One gets more insight looking at the axially symmetric trajectory (see Fig. \ref{Fig2}).  Here, we observe the evolution from a spherical potential ($\zeta=0$), whose minimum is found at $\beta=0$, to potentials with well-deformed minima ($\zeta=0.028,0.030$). For intermediate $\zeta$ values one finds a set of potential energy curves which are practically degenerated along the prolate axis in the interval $\beta\in[0,0.4]$. 
These curves show two minima, an  spherical and a prolate deformed one, and can be found only in a very reduced range of values for the control parameter ($0.0251\leq\zeta\leq0.0255$). In particular, for $\zeta=0.02547$ the spherical and the prolate deformed minima are degenerated and this condition defines precisely the critical point of a first order phase transition where the order parameter is the deformation $\beta$. We denote $V_{crit.}(\beta,\gamma)$ the corresponding critical potential.
The narrowness of the mentioned $\zeta$  interval strongly contrasts  with the integer character of the control parameter in real nuclei. \\ \indent
To compare with the experimental data one should consider the dynamical degrees of freedom, i.e., substitute $V_{crit.}(\beta,\gamma)$ in the Bohr Hamiltonian and solve the Schr\"odinger equation $H_{coll}\Phi = E \Phi$ to find energies and wave functions to calculate transition probabilities. Comparison of the theoretical predictions with the experimental data should help to identify the hypothetical critical point nuclei. This program has been conducted by Iachello \cite{IACHPRL01} in a very elegant and clever way approximating $V_{crit.}(\beta,\gamma)$ in a convenient manner. He assumed that the potential was separable in the variables $(\beta,\gamma)$ and approximated the $\beta$ dependence by a square well in the interval  $\beta\in[0,\beta_{M}]$ and the $\gamma$ one by a harmonic oscillator.  
With these approximations one obtains analytical solutions to the Bohr Hamiltonian which describe the critical point of a phase transition from a spherical to an axially symmetric (prolate) shape and are associated to the so-called X(5) symmetry \cite{IACHPRL01}. \\ \indent
\begin{figure} [t]
\centering
\includegraphics[angle=0,scale=0.5]{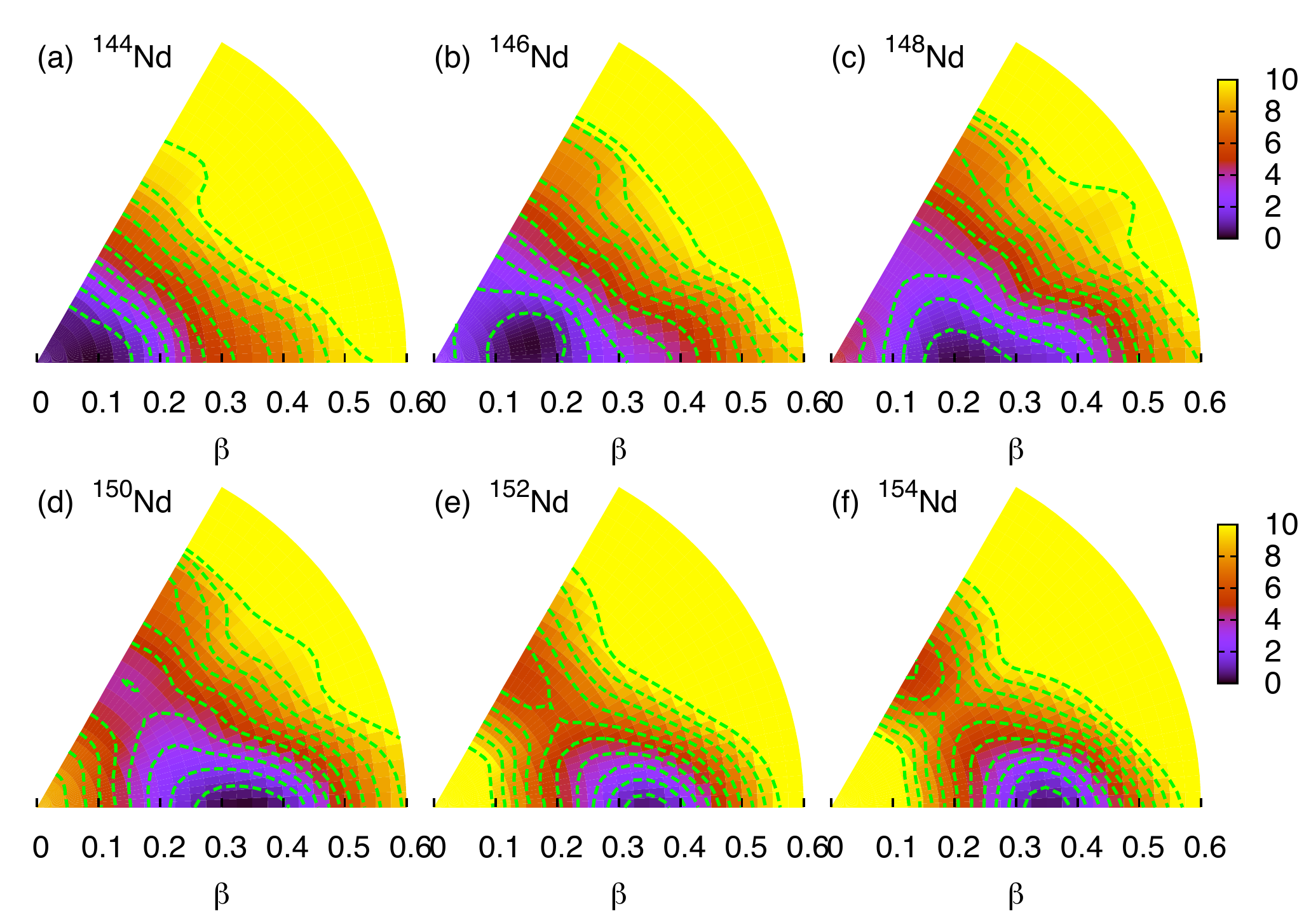} 
\caption{(a)-(f) Particle number projected potential energy surfaces in the ($\beta,\gamma$) plane (Eq. \ref{PNP-PES}) calculated with the Gogny D1S force for $^{144-154}_{60}$Nd. Contour lines represent an step of 1 MeV and the energy is set to zero at the minimum of each surface.}
\label{Fig3}
\end{figure}

Let us now  analyze the PES of the isotopes $^{144-154}$Nd calculated with the Gogny force to see how much they resemble those of Fig. \ref{Fig1}. Since the potential of Eq.~\ref{bohr} preserves the number of particles and the Bohr Hamiltonian recovers the rotational invariance, it is obvious that at the microscopic level the potential $V(\beta,\gamma)$ of Eq.~\ref{bohr} is given by $E^{N,Z}(\vec{q})$, see  Eq.~\ref{PNP-PES}, with $\vec{q}=(\beta,\gamma)$. Furthermore collective hamiltonians similar to the Bohr Hamiltonian can be obtained from the Hill-Wheeler equation considering the Gaussian overlap approximation and also from adiabatic time-dependent Hartree-Fock approximations \cite{RINGSHUCK,K.NPA.74}. Therefore, the diagonalization of the Bohr Hamiltonian is comparable to the solution of a Hill-Wheeler equation.
In Fig.~\ref{Fig3} we show the PES calculated using Eq.~\ref{PNP-PES}, here we observe the shape transition from a practically spherical nucleus ($^{144}$Nd) to well- prolate deformed nuclei ($^{152-154}$Nd). In these limits we appreciate a similarity with the potential surfaces given by the geometrical model (panel \ref{Fig1}(b) in the first case and panel \ref{Fig1}(f) in the deformed one). \\ \indent
However, we also notice several differences between the surfaces given by the microscopic and the geometrical model.
Thus the microscopic ones present more substructure due to the underlying shell structure which is ignored in the collective model. In addition, there is no continuous control parameter in the isotopic chain. Nevertheless, the most important difference is the role played by the triaxial degree of freedom in the calculations with the Gogny force for the transitional isotopes $^{146-150}$Nd. In these cases, we notice that the surfaces soften not only in the axial prolate direction, as in Fig.~(\ref{Fig1}), but also in the $\gamma$ direction.  Finally,  the $^{152-154}$Nd nuclei show symmetric minima isolated from the rest of the $(\beta,\gamma)$ plane with high barriers.\\ \indent
\begin{figure} [t]
\centering
\includegraphics[angle=0,scale=0.5]{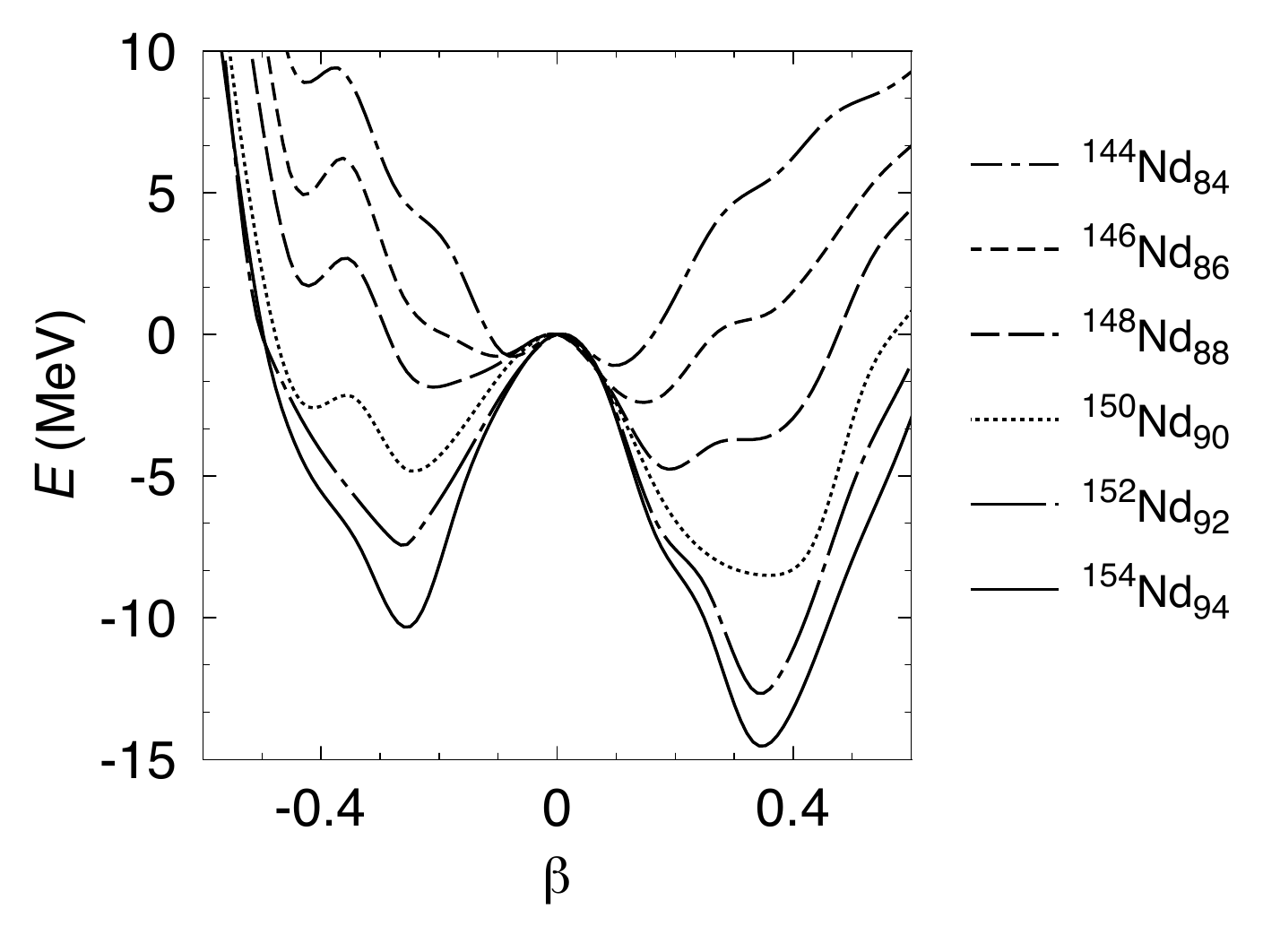} 
\caption{Particle number projected potential energy surfaces along the axial trajectory $\gamma=0^{\circ},60^{\circ}$ calculated with the Gogny D1S force for $^{144-154}_{     60}$Nd. The zero of the energy is chosen at $\beta=0$. }
\label{Fig4}
\end{figure}
In Fig.~\ref{Fig4} we show the potential energy curves along the axial deformation. Here, we observe that prolate and oblate deformed minima are obtained for all nuclei with a maximum at $\beta=0$, in contrast to the result shown in fig. (\ref{Fig2}). The deformation corresponding to those minima is displaced to higher values with increasing  mass number. Furthermore, both the energy difference between the prolate and oblate minima, and also the value of the barrier between them, are larger with larger number of neutrons in the system. Nevertheless, it is important to note that in the $^{144-150}$Nd isotopes, the oblate minima are  saddle points in the $(\beta,\gamma)$ plane, while for the $^{152-154}$Nd nuclei they are real minima.   \\ \indent

The study of the particle number projected PES only gives a qualitative picture of the shape transition. To obtain spectroscopic information and to include configuration mixing, we perform GCM calculations along the axial quadrupole degree of freedom with particle number and angular momentum projected wave functions. As we have already mentioned in the previous paragraph, the triaxial effects may play an important role in the transitional nuclei. However, configuration mixing calculations that include the full triaxial angular momentum projection are not feasible with the current computational capabilities. Therefore, a significant deviation of the theoretical prediction from the experimental data only for
the soft triaxial nuclei could point out to the  relevance of this  degree of freedom. \\ \indent
\begin{figure} [t]
\centering
\includegraphics[angle=0,scale=0.6]{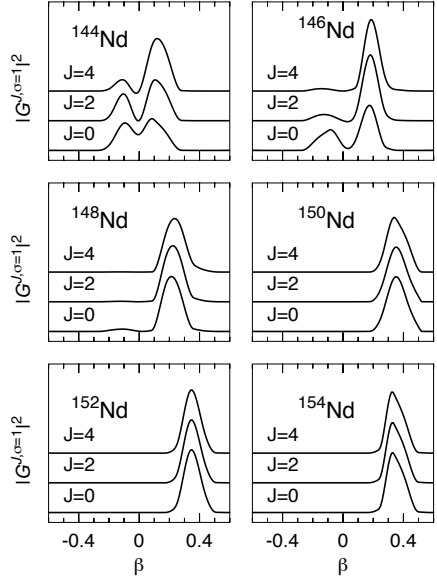} 
\caption{Collective wave functions  along the axial quadrupole deformation ($\beta$) for $J=0$, $J=2$ and $J=4$, calculated for $^{144-154}$Nd isotopes. $|G(\beta)|^2$ represents the probability to find the deformation $\beta$ in the collective wave functions, see also \cite{RODGUZNPA02}}.
\label{Fig5}
\end{figure}
In Fig.~\ref{Fig5} we analyze  the effects  of the angular momentum projection and the GCM on the ground and first excite states  ($J^{\pi}=0^{+},2^{+},4^{+}$).  The  collective wave functions (Eq. \ref{GCMstate}) of the ground states of the  $^{144-146}$Nd isotopes exhibit a noticeable configuration mixing of oblate and prolate shapes, the $^{146}$Nd nucleus being more deformed than the $^{144}$Nd one. The prolate-oblate mixing tends to disappear with increasing value of the angular momentum where clearly only prolate mixing is obtained. On the other hand, concerning the $^{148-154}$Nd nuclei, all the collective wave functions distinctly peak at positive values of $\beta$ deformation for all values of angular momentum. In addition, with increasing values of the mass number these collective wave functions get narrower, and their minima are located at larger values of $\beta$. If triaxial shapes were allowed
in the configuration mixing calculations we will expect larger mixing in the nuclei with softer
surfaces. \\ \indent   
\begin{figure} [t]
\centering
\includegraphics[angle=0,scale=0.45]{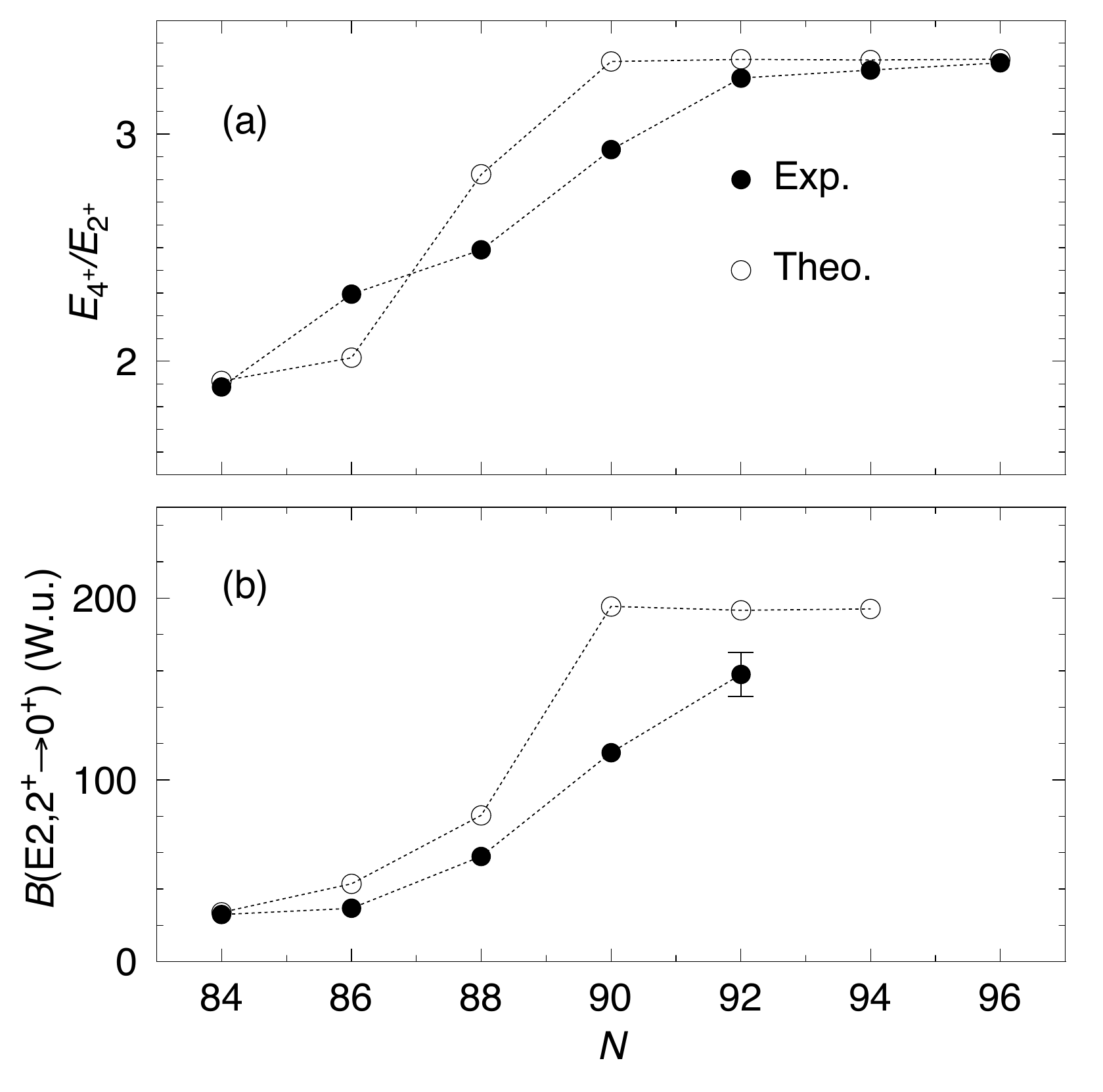} 
\caption{(a) Ratio between the energies of the first $4^{+}$, $2^{+}$ excited states, (b) reduced quadrupole transition probability between the first $2^{+}$, ground states for the $^{144-154}$Nd isotopes. Dots and circles correspond to experimental --- extracted from Refs. \cite{NUCDATA,KRUPRL02} --- and theoretical values, respectively.} 
\label{Fig6}
\end{figure}
One of the best signatures of a shape  transition is the behavior of the ratio between the energies of the first $4^{+}$ and $2^{+}$ states ($R_{4/2}=E(4^{+})/E(2^{+})$) along the isotopic chain. This ratio varies from the value which correspond to vibrations around a spherical shape ($R_{4/2}=2$) to the characteristic value for excitations of a well-deformed rotor ($R_{4/2}=3.33$). Fig.~\ref{Fig6}(a) shows that these limits are experimentally fulfilled in the $^{144}$Nd and $^{152-156}$Nd isotopes, respectively, and also that there is a smooth transition between them.  
Furthermore, we observe a good agreement between theory and experiment, specially in the vibrational and rotational limits. However, we also notice that the transition is more abrupt in the theoretical predictions than in the experimental results and only the $^{148}$Nd isotope could be considered as a transitional nucleus in the calculations. Similar conclusions are extracted from the behavior of the reduced quadrupole transition probabilities  ($B(E2,2^{+}\rightarrow0^{+})$) shown in Fig.~\ref{Fig6}(b). Experimentally and theoretically an increase of the $B(E2)$'s with increasing neutron number is observed  as we are moving further away from the $N=82$ shell closure. Again, the theoretical results display
a sharper jump from the weak to the strong collective behavior with $^{148}$Nd in an intermediate 
position.\\ 
\indent
\begin{figure} [t]
\centering
\includegraphics[angle=0,scale=0.6]{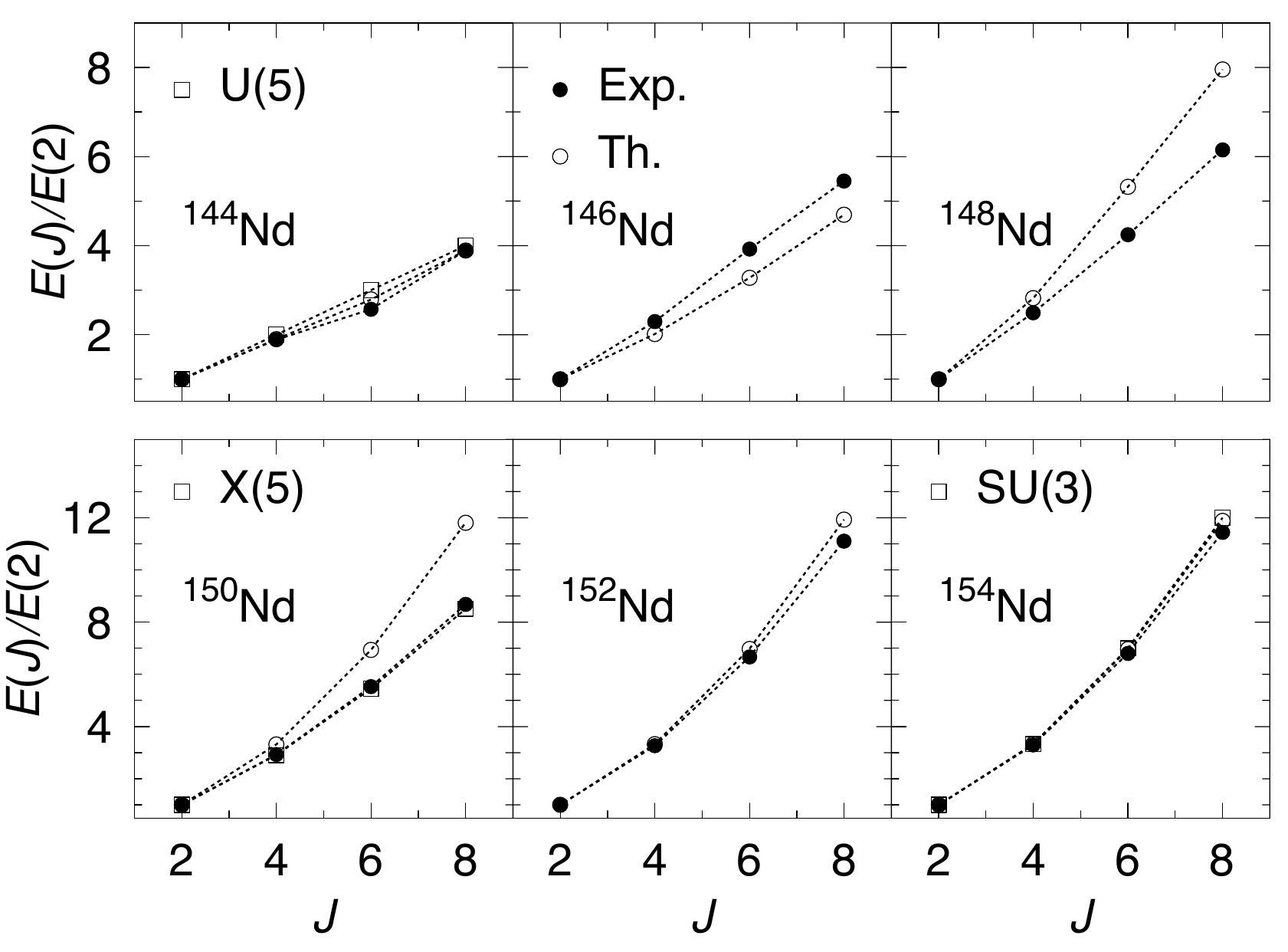} 
\caption{Normalized excitation energies $E(J^{+}_{1})/E(2^{+}_{1})$ for the $^{144-154}$Nd isotopes. Dots and circles correspond to experimental and theoretical values, respectively. In addition, U(5), X(5) and SU(3) predictions (boxes) are shown in comparison to the data for $^{144}$Nd, $^{150}$Nd and $^{154}$Nd nuclei. Notice the different scale in the upper and lower panels.}
\label{Fig7}
\end{figure}
We also compare our theoretical results with the available experimental data \cite{NUCDATA,KRUPRL02} for higher excited states in Fig.~\ref{Fig7}. The experimental energy spectra in the vibrator ($^{144}$Nd) and rotor ($^{152-154}$Nd) limits are nicely described with the present calculations. Additionally, we plot in Fig. \ref{Fig7} the predictions given by U(5) and SU(3) models for those cases and an excellent matching between both the symmetries and the experimental and theoretical results are observed. In the first case, the law $E(J)\sim J$ is obeyed while in the latter $E(J)\sim J(J+1)$ is fulfilled. However, deviations from the experimental trends are found in $^{146-150}$Nd.  It is precisely in these nuclei where the triaxial effects --- not included in these calculations --- are expected to be more important (see Fig. \ref{Fig3}).\\ \indent
In Fig. \ref{Fig7} the predictions of the X(5) symmetry \cite{IACHPRL01} are also shown. There, a good agreement with the experimental data for the $^{150}$Nd isotope is found and, from this kind of comparisons, it has been inferred that this nucleus corresponds to the experimental critical point \cite{KRUPRL02}. 
We have already mentioned that in our description it is the nucleus $^{148}$Nd  that looks more transitional. As a matter of fact, if we compare the X(5) results with the theoretical predictions for $^{148}$Nd, see Fig. \ref{Fig8}, we find striking similarities in energies and transition probabilities suggesting that the theoretical results for the nucleus $^{148}$Nd display the X(5) symmetry.
\begin{figure} [t]
\centering
\includegraphics[angle=0,scale=.5]{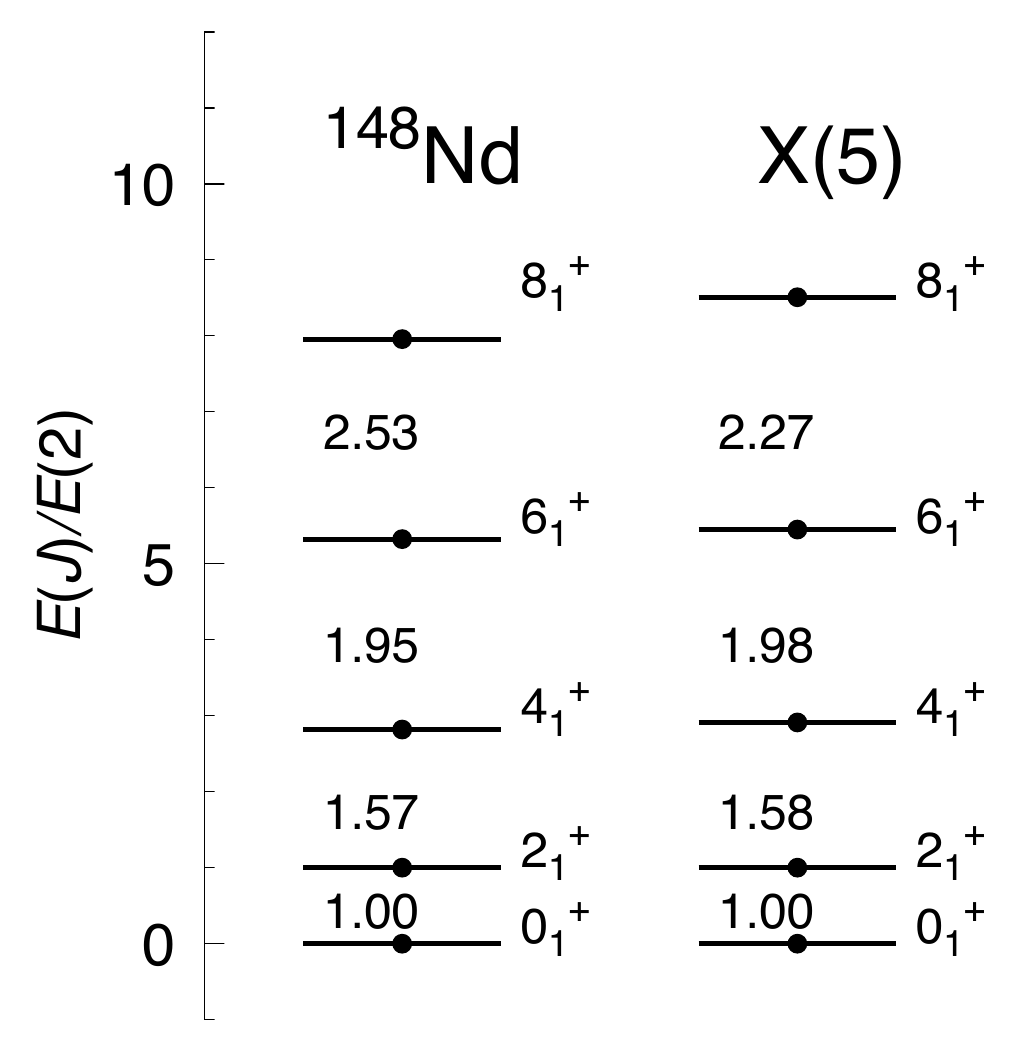}
\caption{Normalized excitation energies and transition probabilities for the $^{148}$Nd isotope calculated with the GCM method and the X(5) predictions.}
\label{Fig8}
\end{figure}
As mentioned at the beginning there have been also configuration mixing calculations 
performed in the framework
of the relativistic model by Nik\u{s}i\'c and collaborators \cite{NVLR.PRL.07}. These authors obtain a good agreement with the experimental
results for the nucleus $^{150}$Nd and we can  affirm that the relativistic models also reproduce 
the X(5) predictions. At this point we are confronted with a kind of paradox, the {\it must condition} of Iachello to  
obtain the X(5) symmetry spectrum is to impose that in the critical point nuclei the spherical and the deformed minima must coexist and be degenerated in order to obtain a first order shape phase transition. However, our 
calculations and the relativistic ones are able to produce  X(5) results with potentials   very much different from the one of Iachello, see Fig.~\ref{Fig4} of the present work and  Fig. 1 of Ref.~\cite{NVLR.PRL.07}. 
It is important to realize that the potentials used in the relativistic  calculations as well as in ours do not satisfy the mentioned {\it must condition} to identify a critical point nuclei. Hence, it seems that spectra and transition probabilities of the X(5) type are not sufficient for the identification of a 
critical point nucleus, at least not for the type of phase transition described in Ref.~\cite{IACHPRL01}.
One could think that the relativistic potential of Ref. \cite{NVLR.PRL.07} and ours look similar and that it is by chance that the calculations reproduce the X(5) results. However, this is not the case because Nik\u{s}i\'c et al.  only consider the prolate part of the potential in their configuration mixing calculations and neglect the mixing with oblate configurations. This is not the case
in our calculations and as a matter of
fact we have performed calculations both with and without the oblate part and we find large differences. As one can see in Fig. \ref{Fig9}, where  for the nuclei  $^{146}$Nd and $^{148}$Nd we plot the spectra with and without consideration of the oblate part of the potential. In particular for the nucleus  $^{146}$Nd we find that in the complete calculations we are close to the vibrational limit and in the prolate restricted ones close to the rotational limit.
\begin{figure} [t]
\centering
\includegraphics[angle=0,scale=.7]{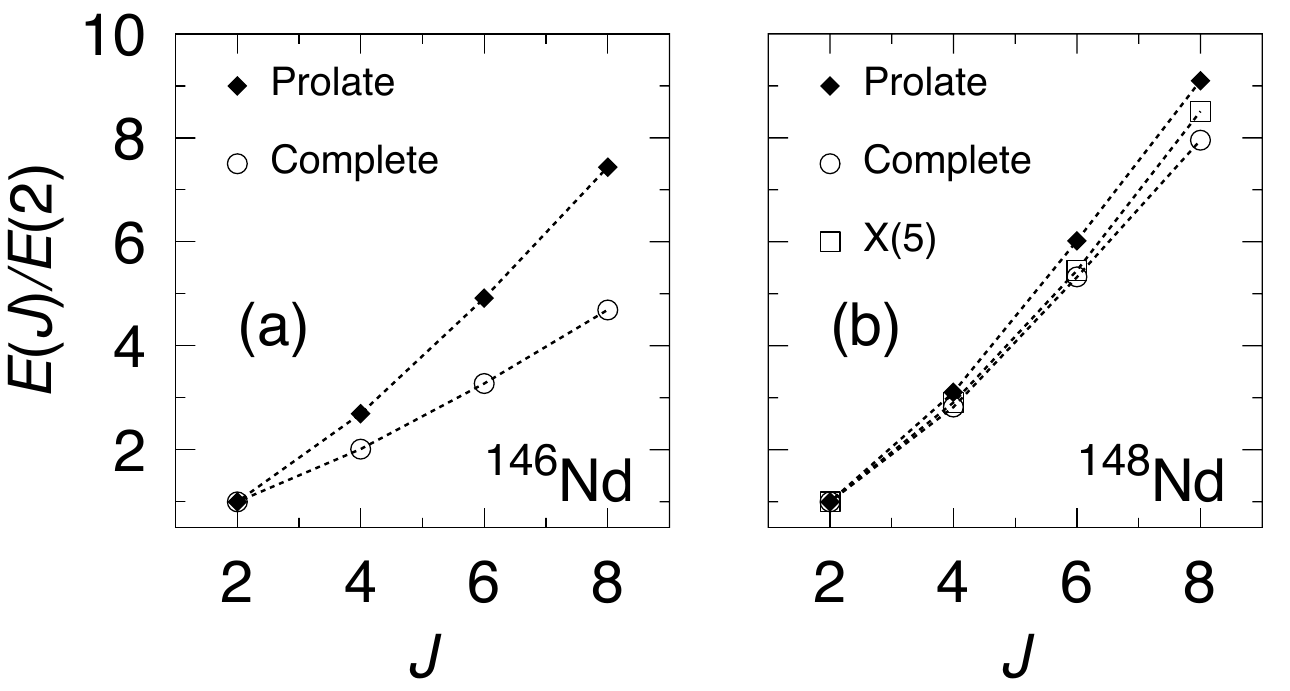}
\caption{Normalized excitation energies $E(J^{+}_{1})/E(2^{+}_{1})$ for (a) $^{146}$Nd and (b) $^{148}$Nd isotopes calculated with the GCM method using only prolate (diamonds) and prolate and oblate configurations (circles) with the Gogny force. Furthermore, the experimental data (dots) and the X(5) predictions (boxes) are also represented.}
\label{Fig9}
\end{figure}
A relevant aspect of our calculations is that they reveal the importance of triaxial shapes in the transitional nuclei $^{146-148-150}$Nd and the potential role that they should  play if considered in the calculations. One has further to consider that triaxial angular momentum projection of the PES of Fig. \ref{Fig3} will make them shallower, see Ref. \cite{Yang_Sun}, and the subsequent configuration mixing of triaxial shapes even more relevant.  Another point which 
in our opinion
has not received enough attention in earlier works is the presence of negative parity states with $J^{\pi}=3^{-}$ and $J^{\pi}=1^{-}$ experimentally observed  in the low-lying energy spectra for the $^{146-150}$Nd  \cite{NUCDATA,IW.PRL.93} and Samarium isotopes \cite{NUCDATA}. These data indicate the relevance of the octupole degree of freedom in this region.
Though our present parity conserving calculations ignore this degree of freedom, earlier calculations with the Gogny force which had considered it, have shown \cite{GARROTEPRL98} that the Nd isotopes are rather soft against octupole deformation. Furthermore, these calculations were able to reproduce these negative parity states \cite{GARROTEPRL98} up to high angular momentum. 

In conclusion we have studied the spherical to prolate deformed shape transition in the Neodymium isotopes ($^{144-154}$Nd) with state-of-the-art beyond mean field calculations with the Gogny force. 
We have applied  the generator coordinate method with particle number and angular momentum projected axial wave functions to this problem. We have found a good agreement with the experimental data, specially in the vibrational and rotational limits, although the calculations are not able to reproduce quantitatively the transitional region. As a matter of fact, the transition is theoretically more abrupt than in the experiments. 
Concerning the transitional region we would like to underline the following points:  First,  there are several intrinsic potentials which display spectra and transition probabilities corresponding  to the X(5) symmetry and which do not satisfy the condition, given in Ref.~\cite{IACHPRL01}, to identify them with the critical point of a first order shape phase transition. Second, the simplistic view point of considering in a phase transition all degrees of freedom frozen but the axial quadrupole moment is not supported by microscopic
calculations. On the contrary, these calculations seem to support a general relaxation of the most relevant degrees of freedom. Third and last, the consideration of the beta deformation as the order parameter of a hypothetical shape phase transition is not well founded.  All these points question the interpretation
of nuclear shape changes as nuclear shape phase transitions.
The final interesting task for the future is to see whether the inclusion of additional degrees of freedom (specially the triaxial one) leads us to a quantitative description of the transition region.

This work has been supported in part by the DGI, Ministerio de Ciencia y Tecnologia, Spain, 
under Project FIS2004-06697.

\end{document}